\documentclass[twocolumn,secnumarabic,amssymb, amsmath, nofootinbib,tightenlines,
nobibnotes, aps, prl,epsfig]{revtex4}
\usepackage{graphicx}
\usepackage{dcolumn}
\usepackage{bm}
\begin{document}
\preprint{APS/123-QED}
\title{Effects of heavy quarks in the dipole cross section with the KLN model}

\author{G.R.Boroun}%
 \email{boroun@razi.ac.ir}

 \affiliation{Department of physics, Razi University, Kermanshah
67149, Iran}
\date{\today}
\begin{abstract}
The dipole cross-section behavior for protons and nuclei is
analyzed with and without considering the heavy quark masses in
the Bjorken variable $x$ using the Kharzeev-Levin-Nardi (KLN)
model of low $x$ gluon distributions. The Color Glass Condensate
(CGC) effects in the color dipole model are influenced by the
heavy quark masses in the Bjorken variable $x$ at $Q^2<Q_{s}^2$.
These non-linear saturation effects will be
observable in the LHeC and EIC at very small $x$ values.\\

\end{abstract}
\maketitle
\subsection{I. Introduction}
 The saturation model is particularly intriguing  when describing deep
 inelastic scattering (DIS) data at low $x$, as the saturation of the growth of
gluon densities in hadrons and nuclei is evident in this region
\cite{Ref1, Ref2, Ref3, Ref4}. A simple model by Kharzeev, Levin,
and Nardi (KLN) in Ref.\cite{Ref5} was proposed to explain
saturation physics for the unintegrated gluon distribution
function. This model only provides an Ansatz for the gluon
distribution function, which is suitable for color dipole models
modified into the gluon density. Hadron physics in new
accelerators, such as LHeC and EIC,  relies on the saturation
scale, $Q_{s}^2$, in the KLN model which is interpreted
as evidence for parton recombination and saturation.\\
A new regime of quantum chromodynamics (QCD) is characterized by
high gluon densities at $Q^2<Q_{s}^2$. In this regime, gluon
densities at small $x$, in a hadron wavefunction, represent the
Color Glass Condensate (CGC) \cite{Ref6, Ref7}. The boundary
between dense and dilute gluonic systems is related to the
presence of large quark masses in the Bjorken variable $x$. In
this paper, we demonstrate that the non-linear saturation dynamics
incorporated into the CGC model are observable when using the
Bjorken scaling for the proton color dipole cross section at very
small $x$, but are eliminated when the Bjorken variable  depends
on charm and bottom quark masses. For light and heavy nuclei, the
non-linear saturation dynamics are apparent at very small $x$ when
considering the modified Bjorken variable $x$.

\subsection{II. Dipole cross section}
The proton structure function $F_{2}$ is related to the
 $\gamma^{*}p$ cross section as
 \begin{eqnarray}
F_{2}(x,Q^2)=\frac{Q^2}{4\pi^2\alpha_{em}}\bigg{(}\sigma^{\gamma^{*}p}_{T}(x,Q^2)+\sigma^{\gamma^{*}p}_{L}(x,Q^2)\bigg{)}.
\end{eqnarray}
In the color dipole picture (CDP), the scattering between the
virtual photon $\gamma^{*}$ and the proton is seen in the
following form
\begin{eqnarray}
\sigma_{L,T}^{\gamma^{*}p}(x,Q^{2})=\int dz d^{2}\mathbf{r}
|\Psi_{L,T}(\mathbf{r},z,Q^{2})|^{2}\sigma_{\mathrm{dip}}(\widetilde{{x}},\mathbf{r}).
\end{eqnarray}
The virtual photon dissociates into a quark-antiquark pair (a
$q\overline{q}$ dipole) which interacts with the color fields in
the proton. The imaginary part of the $(q\overline{q})p$ forward
scattering amplitude is defined by the dipole cross-section,
$\sigma_{\mathrm{dip}}({x},r)$. The variables $r$ and $z$ are
defined, with respect to the photon
 momentum, representing the transverse dipole size and the longitudinal momentum
 fraction respectively \cite{Ref1, Ref2, Ref3, Ref4}. $\Psi_{L,T}$ is the photon wave function with respect to
the photon polarization and depends on the mass of the quarks in
the $q\overline{q}$ dipole. Therefore, the Bjorken variable $x$ is
modified by the following form \cite{Ref8}
\begin{eqnarray}
x{\rightarrow}\widetilde{x}_{f}=x\bigg{(}1+\frac{4m_{f}^{2}}{Q^2}\bigg{)},
\end{eqnarray}
where $m_{f}$ is the quark mass.\\
The dipole cross-section in the Golec-Biernat and
W$\mathrm{\ddot{u}}$sthoff (GBW) model \cite{Ref4}, depends on the
dipole size $r$ and the saturation scale $Q_{\mathrm{sat}}(x)$ as
follows
\begin{eqnarray}
\sigma_{\mathrm{dip}}(\widetilde{x},\mathbf{r})=\sigma_{0}\bigg{\{}1-
\exp\bigg{(}-r^2Q_{\mathrm{sat}}^2(\widetilde{x})/4\bigg{)}
\bigg{\}},
\end{eqnarray}
where the saturation scale is parametrized as
$Q_{\mathrm{sat}}^2(\widetilde{x})=Q_{0}^{2}(x_{0}/\widetilde{x})^\lambda$.
The dipole cross section was improved by taking into account the
evolution of the gluon density in Ref.\cite{Ref9}. The modified
dipole cross section in the Bartels, Golec-Biernat and Kowalski
(BGK) model \cite{Ref9} reads
\begin{eqnarray}
\sigma_{\mathrm{dip}}(\widetilde{x},\mathbf{r})=\sigma_{0}\bigg{\{}1-
\exp\bigg{(}-
\frac{\pi^2r^2\alpha_{s}(\mu^2)xg(\widetilde{x},\mu^2)}{3\sigma_{0}}
\bigg{)} \bigg{\}},
\end{eqnarray}
where the evolution scale $\mu^2$ is connected to the size of the
dipole by $\mu^2=\frac{C}{r^2}+\mu^2_{0}$.\\
In the following we shall use the KLN Ansatz \cite{Ref10} in the
dipole cross section in the following form
\begin{eqnarray}
\sigma_{\mathrm{dip}}(\widetilde{x},\mathbf{r})&=&\sigma_{0}\bigg{\{}1-
\exp\bigg{(}-
\frac{\pi^2r^2\alpha_{s}(\mu^2)}{3\sigma_{0}}\bigg{[}\nonumber\\
&&\frac{K_{0}S}{\alpha_{s}(Q_{s}^{2})}\mu^2(1-x)^{D}\Theta(Q_{s}^{2}-\mu^{2})\nonumber\\
&&+\frac{K_{0}S}{\alpha_{s}(Q_{s}^{2})}Q_{s}^{2}(1-x)^{D}\Theta(\mu^{2}-Q_{s}^{2})
\bigg{]} \bigg{)} \bigg{\}},
\end{eqnarray}
where $S$ is the area of the target and $K$ is a constant
parameter obtained from the momentum sum rule. The nuclear dipole
cross section $\sigma^{A}_{\mathrm{dip}}$ with the KLN
distribution can accurately predict data from deep inelastic
scattering in electron-ion collisions (EICs)
\cite{Ref11,Ref12,Ref13}. For a nuclear target with the mass
number A, the nuclear dipole cross section remains the same for
the change $xg(x,Q^2){\rightarrow}xg^{A}(x,Q^2)$ as
\begin{eqnarray}
\sigma^{A}_{\mathrm{dip}}(\widetilde{x},\mathbf{r})&=&\sigma^{A}_{0}\bigg{\{}1-
\exp\bigg{(}-
\frac{\pi^2r^2\alpha_{s}(\mu^2)}{3\sigma^{A}_{0}}\bigg{[}\nonumber\\
&&\frac{K_{0}S_{A}}{\alpha_{s}(Q_{sA}^{2})}\mu^2(1-x)^{D}\Theta(Q_{sA}^{2}-\mu^{2})\nonumber\\
&&+\frac{K_{0}S_{A}}{\alpha_{s}(Q_{sA}^{2})}Q_{sA}^{2}(1-x)^{D}\Theta(\mu^{2}-Q_{sA}^{2})
\bigg{]} \bigg{)} \bigg{\}},~~~
\end{eqnarray}
where the replacements are $S^{A}=A^{2/3}S$,
$\sigma_{0}^{A}=A^{2/3}\sigma_{0}$ and
$Q^{2}_{sA}=A^{1/3}Q^{2}_{s}$ \cite{Ref10, Ref14}.\\
By considering the individual quark flavor pairs in the QCD dipole
picture, the rescaling Bjorken variable $x$ (i.e., Eq.(3)) is
modified from light quark pairs to heavy quark pairs (i.e.,
$c\overline{c}$ and $b\overline{b}$ pairs). Indeed the rescaling
variable $\widetilde{x}_{f}=x{(}1+\frac{4m_{f}^{2}}{\mu^2}{)}$ is
introduced \cite{Ref15} to extend the saturation model to the low
$\mu^2$ (large $r$) region including the photoproduction limit.
The dipole cross sections for nucleons and nuclei (i.e., Eqs.(6)
and (7)) in the KLN model satisfy the condition of
$Q^2>Q^{2}_{sA}$ (for proton A=1) when the rescaling Bjorken
variable $x$ is considered. Without the rescaling variable, the
dipole cross section behavior will deviate from the GBW model when
it satisfies the condition $Q^2<Q^{2}_{sA}$ at large $r$. This
deviation is supported by the CGC model \cite{Ref16, Ref6, Ref7}.
In the next section we will consider the behavior of the dipole
cross section with and without the rescaling effects for nucleons
and nuclei. Indeed, the rescaling effect is one of the ingredients
used in the general-mass variable
flavor number schemes (GM-VFNS) \cite{Ref17}.\\

\subsection{II. Results and Conclusions}

The fixed parameters are summarized in Table I for the quark
masses of Fits 0, 1 and 2 with the quark masses
$m_{l}=0.14~\mathrm{GeV}$, $m_{c}=1.4~\mathrm{GeV}$ and
$m_{b}=4.6~\mathrm{GeV}$ respectively, as shown in
Ref.\cite{Ref5}. We have calculated the $r$-dependence of the
ratio $\sigma_{\mathrm{dip}}/\sigma_{0}$ (i.e., Eq.(6)) based on
the KLN model for $x=10^{-6}$ and $10^{-3}$. In Figs.1 and 2, the
results of the KLN model are compared with the GBW model,
considering the charm and bottom masses in the rescaling variable
$x$, respectively. The coefficients in the CDP are based on fits 1
and 2 in Table I, as shown in Figs.1 and 2 respectively. The
rescaling variable $x$ (i.e., Eq.(3)) is applied to the GBW and
KLN models as we observe that the results are comparable in these
figures (i.e., Figs.1 and 2). In wide ranges of $x$ and $r$, the
KLN results fall within the domain $\mu^2>Q_{s}^{2}$ when we
consider the rescaling variable with charm and bottom masses.
Indeed, the
rescaling variable ensures the GM-VFNS in the KLN model.\\
\begin{table}
\centering \caption{The fixed parameters of the color dipole model
from the fit results in  Ref.\cite{Ref8}.
  }\label{table:table1}
\begin{minipage}{\linewidth}
\renewcommand{\thefootnote}{\thempfootnote}
\centering
\begin{tabular}{|l|c|c|c|c|c|} \hline\noalign{\smallskip}
Fit & C & $\mu_{0}^{2}~[\mathrm{GeV}^2]$ & $\sigma_{0}~[\mathrm{mb}]$  & $\lambda$ & $x_{0}{\times}10^{-4}$\\
\hline\noalign{\smallskip}
0 & 0.29 & 1.85 & 23.58 & 0.270 & 2.24  \\
1 & 0.29 & 1.85 & 27.32 & 0.248 & 0.42  \\
2 & 0.27 & 1.74 & 27.43 & 0.248 & 0.40  \\
\hline\noalign{\smallskip}
\end{tabular}
\end{minipage}
\end{table}
The ratio $\sigma_{\mathrm{dip}}/\sigma_{0}$ in the KLN model is
calculated  without considering the charm and bottom masses in
Figs.3 and 4. Without the rescaling variable $x$, the hard
saturation momentum $Q_{s}(x)$ grows rapidly as $x$ decreases at
large values of $r$. In Figs.3 and 4, we observe that the
condition $\mu^2>Q^{2}_{s}$ is valid at small $r$, and the
condition $\mu^2<Q^{2}_{s}$ at large $r$ is valid. For
$\mu^2<Q^{2}_{s}$ at large $r$ the saturation becomes visible. The
deviations of the ratio $\sigma_{\mathrm{dip}}/\sigma_{0}$ from a
straight line at low $x$ values (specifically at $x=10^{-6}$),
indicate the significance of non-linear effects in the KLN model
without the rescaling variable. In reality, saturation effects are
only noticeable at very small $x$ ($x=10^{-6}$) and for large
values of $r$. This depletion  in the ratio is referred to as
shadowing. The depletion point increases towards larger $r$ with
an increase in the production of heavy quarks from $c\overline{c}$
to $b\overline{b}$ in the CDP. For the production of
$c\overline{c}$ in the CDP this depletion point is at $r>0.06$ in
Fig.3 and at $r>0.1$ for the
production of $b\overline{b}$ in Fig.4.\\
\begin{figure}
\centering
\includegraphics[width=0.55\textwidth]{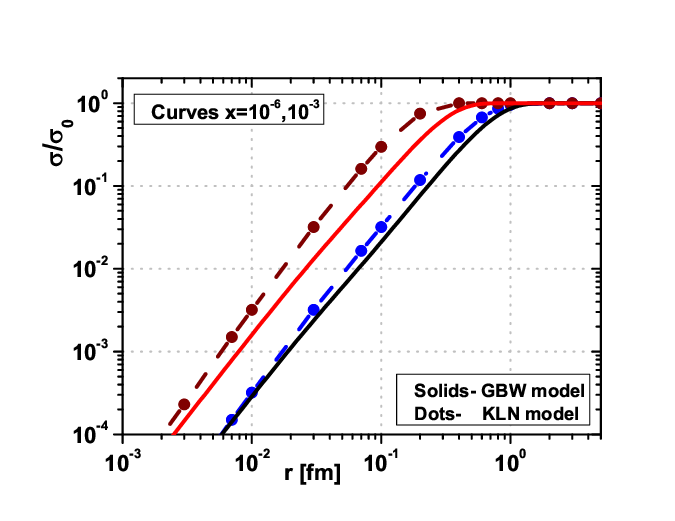}
\caption{The extracted ratio  $\sigma_{\mathrm{dip}}/\sigma_{0}$
as a function of $r$ for $x=10^{-6}$ (brown curve) and $x=10^{-3}$
(blue curve) (curves are from left to right, respectively) using
the KLN model, Eq.(6), compared with the GBW model (solid curves),
Eq.(4). In these results the charm effect in the rescaling Bjorken
 variable $x$ (i.e., Eq.(3)) is considered. The coefficients
  are based on the results from Fit 1 in Table I.}
\end{figure}
\begin{figure}
\centering
\includegraphics[width=0.55\textwidth]{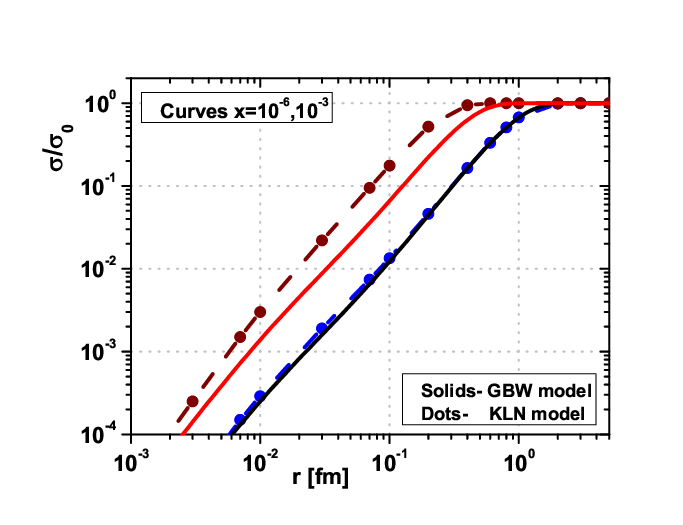}
\caption{The extracted ratio  $\sigma_{\mathrm{dip}}/\sigma_{0}$
as a function of $r$ for $x=10^{-6}$ (brown curve) and $x=10^{-3}$
(blue curve) (curves are from left to right, respectively) using
the KLN model, Eq.(6), compared with the GBW model (solid curves),
Eq.(4). In these results the bottom effect in the rescaling
Bjorken
 variable $x$ (i.e., Eq.(3)) is considered. The coefficients
  are based on the results from Fit 2 in Table I.}
\end{figure}
\begin{figure}
\centering
\includegraphics[width=0.55\textwidth]{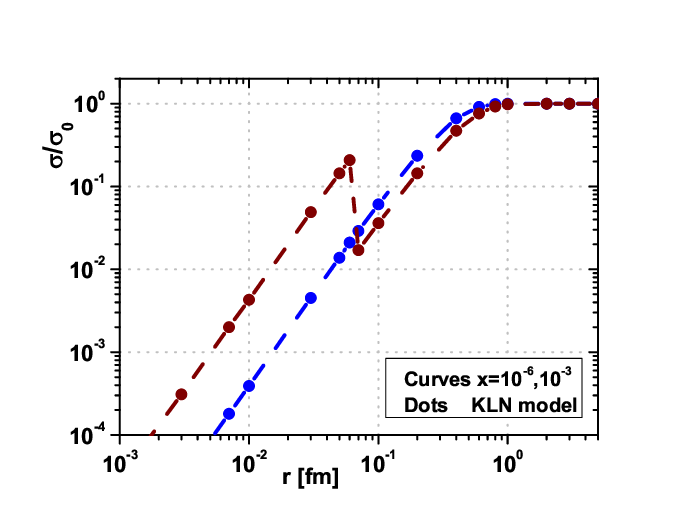}
\caption{The extracted ratio  $\sigma_{\mathrm{dip}}/\sigma_{0}$
as a function $r$ for $x=10^{-6}$ (brown curve) and $x=10^{-3}$
(blue curve) (curves are from left to right, respectively) with
the KLN model, Eq.(6) without the charm effect in the rescaling
Bjorken variable $x$. The coefficients
  are based on the results from Fit 0 in Table I.}
\end{figure}
\begin{figure}
\centering
\includegraphics[width=0.55\textwidth]{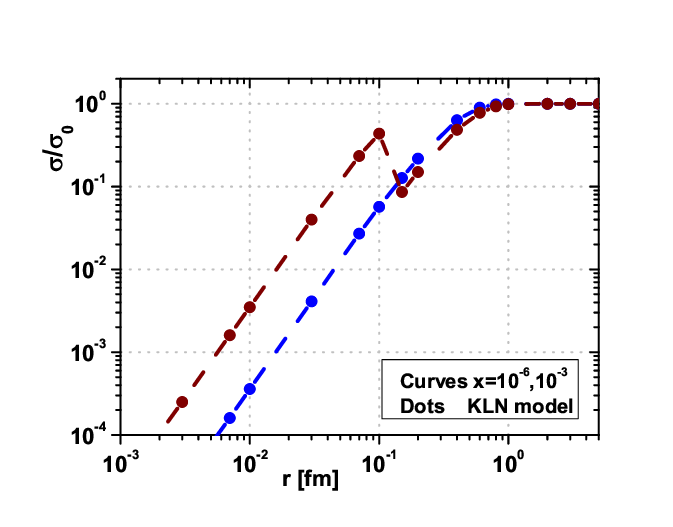}
\caption{The extracted ratio  $\sigma_{\mathrm{dip}}/\sigma_{0}$
as a function $r$ for $x=10^{-6}$ (brown curve) and $x=10^{-3}$
(blue curve) (curves are from left to right, respectively) with
the KLN model, Eq.(6) without the bottom effect in the rescaling
Bjorken variable $x$. The coefficients
  are based on the results from Fit 1 in Table I.}
\end{figure}
It is interesting that the dipole cross sections in Figs.3 and 4
have a property of geometrical scaling, without the rescaling
Bjorken variable $x$. This means they become independent of $x$
and $r$ at large $r$. The KLN criterion is a border between dense
and dilute gluonic systems, which is defined by the geometrical
scaling and is observable with and without the rescaling variable
in the dipole cross sections \cite{Ref18}.\\
In nuclear targets, non-linear effects in electron-nucleus (eA)
processes are evident even when rescaling the Bjorken variable in
the dipole cross sections due to the KLN model. The KLN
prescription for the CGC dynamics is discussed in
Ref.\cite{Ref19}. This saturation refers to the very small-$x$
evolution effects in the dipole cross sections and will be one of
the key physics goals of an Electron-Ion Collider.\\
The ratio $\sigma^{A}_{\mathrm{dip}}/\sigma^{A}_{0}$ for the light
and  heavy nucleus of C-12 and Pb-208 as a function of $r$ with
respect to the rescaling of the Bjorken variable with charm mass
effect for $x=10^{-3}$ and $10^{-6}$ is shown in Figs.5 and 6
respectively.
\begin{figure}
\centering
\includegraphics[width=0.55\textwidth]{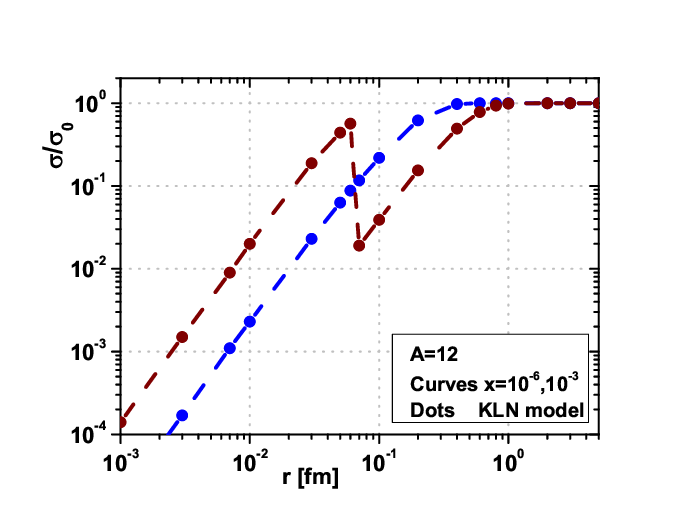}
\caption{The extracted ratio  $\sigma_{\mathrm{dip}}/\sigma_{0}$
for C-12 as a function of $r$ for $x=10^{-6}$ (brown curve) and
$x=10^{-3}$ (blue curve) (curves are from left to right,
respectively) using the KLN model, Eq.(4). In these results the
charm effect in the rescaling Bjorken
 variable $x$ (i.e., Eq.(3)) is considered. The coefficients
  are based on the results from Fit 1 in Table I.}
\end{figure}
\begin{figure}
\centering
\includegraphics[width=0.55\textwidth]{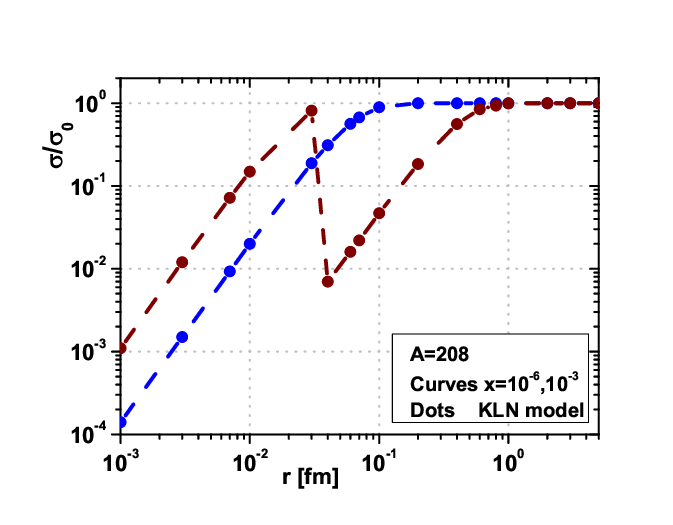}
\caption{The same as Fig.5 for Pb-208.}
\end{figure}
In Figs.7 and 8, the ratio
$\sigma^{A}_{\mathrm{dip}}/\sigma^{A}_{0}$ for the light and heavy
nucleus of C-12 and Pb-208 as a function of $r$ with respect to
the rescaling of the Bjorken variable with bottom mass effect for
$x=10^{-3}$ and $10^{-6}$ is shown respectively.\\
\begin{figure}
\centering
\includegraphics[width=0.55\textwidth]{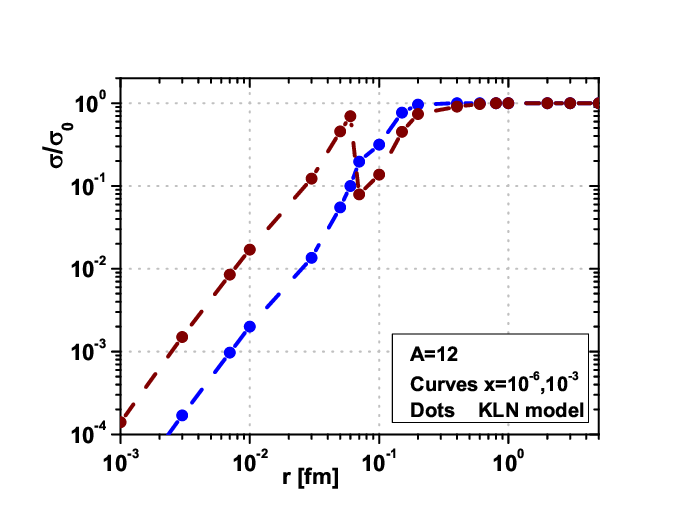}
\caption{The extracted ratio  $\sigma_{\mathrm{dip}}/\sigma_{0}$
for C-12 as a function of $r$ for $x=10^{-6}$ (brown curve) and
$x=10^{-3}$ (blue curve) (curves are from left to right,
respectively) using the KLN model, Eq.(4). In these results the
bottom effect in the rescaling Bjorken
 variable $x$ (i.e., Eq.(3)) is considered. The coefficients
  are based on the results from Fit 2 in Table I.}
\end{figure}
\begin{figure}
\centering
\includegraphics[width=0.55\textwidth]{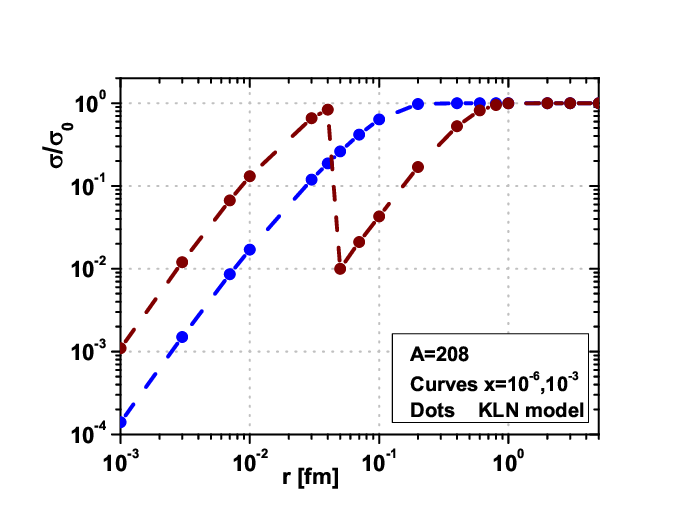}
\caption{The same as Fig.7 for Pb-208.}
\end{figure}
The deviation of the ratios
$\sigma^{A}_{\mathrm{dip}}/\sigma^{A}_{0}$ at very low $x$ (i.e.,
$x=10^{-6}$) in Figs.5-8 shows the importance of non-linear
effects in the KLN model in EICs. Indeed, saturation effects due
to the rescaling variable are noticeable at very low values of $x$
and for large values of the dipole size $r$. This saturation by
the KLN model will be visible in EICs as described in
Ref.\cite{Ref20} by the b-CGC model\footnote{This model
incorporates both the exponential of the two-gluon
exchange and the CGC physics of the saturation of the gluon density.}.\\
The non-linear effects caused by the charm effects in the
rescaling Bjorken variable $x$ at $x=10^{-6}$ are visible in Fig.5
for the light nucleus of C-12 at $r{\gtrsim}0.04$ and in Fig.6 for
the heavy nucleus of Pb-208 at $r{\gtrsim}0.02$ respectively.
Depletions observed in Figs.5 and 6 clearly demonstrate shadowing
effects in the EICs at very low $x$ and large $r$. It is important
to emphasize that the non-linear effects are not visible at small
$x$ (i.e., $x{\approx}10^{-3}$).\\
In Figs.7 and 8 the non-linear effects caused by the bottom
effects in the rescaling Bjorken variable $x$ at $x=10^{-6}$ are
observable for the light nucleus of C-12 at $r{\gtrsim}0.06$ and
 for the heavy nucleus of Pb-208 at $r{\gtrsim}0.04$
respectively. The scale $r$ for the non-linear effects for light
and heavy nuclei increases as the heavy quark mass effect at the
rescaling variable increases from charm to bottom masses. We
observe that, in Figs.6 and 8, depletion of the non-linear effects
for  heavy nuclei is deeper than those for light nuclei in Figs.5
and 7 at very low $x$ and large $r$. We observe that significant
non-linear effects begin to appear at smaller values of $x$ for
heavy nuclei \cite{Ref21, Ref22, Ref23}.\\

In conclusion, we have considered the non-linear saturation
effects in the dipole cross section using the KLN model while also
accounting for  heavy quark masses in the Bjorken variable $x$.
This approach demonstrates the CGC model in the color dipole cross
section for both protons and nuclei at very low values of the
Bjorken variable $x$. The $Q^{2}_{s}$ dependence of
$\sigma_{\mathrm{dip}}$ more accurately determined with and
without consideration of the charm and bottom masses in the dipole
model. We anticipate that the observable effects of non-linear
saturation will be seen in the LHeC and EIC accelerators.\\

\subsection{ACKNOWLEDGMENTS}
G.R.Boroun would like to thank Professor F.S. Navarra for useful
comments and invaluable support.

\end{document}